\documentclass[%
superscriptaddress,
twocolumn,
amsmath,amssymb,
aps,%prl
prl,
]{revtex4-1}
\usepackage{mwe}
\usepackage{bm}
\usepackage{graphicx}
\usepackage[normalem]{ulem}
\usepackage{graphicx}
\usepackage{epstopdf}
\usepackage{hyperref}
\usepackage[normalem]{ulem}
\usepackage{xcolor}
\usepackage{bm}
\usepackage{url}
\usepackage{lipsum}
\usepackage{babel}
\usepackage{float}
\usepackage{dsfont}

\def \g{{\gamma}}
\def \ve{{\varepsilon}}

\def \bs{\boldsymbol}

\newcommand{\s}{\sigma}
\newcommand{\e}{\epsilon}

\newcommand{\D}{\Delta}

\usepackage{bm}

\newcommand{\TSB}{4Hb-$\mathrm{TaS_{2}}$ }	

\begin{document}

\title{Evidence of a two-component order parameter in \TSB in the Little-Parks effect}
%%%%%%%%%%%%%%%%%%%%%%%

\author{Avior Almoalem}
\affiliation{Physics Department, Technion-Israel Institute of Technology, Haifa 32000, Israel.}
\author{Irena Feldman}
\affiliation{Physics Department, Technion-Israel Institute of Technology, Haifa 32000, Israel.}
\author{Michael Shlafman}
\affiliation{Andrew and Erna Viterbi Faculty of Electrical and Computer Engineering, Technion, Haifa 32000, Israel.}
\author{Yuval E. Yaish}
\affiliation{Andrew and Erna Viterbi Faculty of Electrical and Computer Engineering, Technion, Haifa 32000, Israel.}
\author{Mark H. Fischer}
\affiliation{Department of Physics, University of Zurich, Winterthurerstrasse 190, 8057 Zurich, Switzerland}
\author{Michael Moshe}
\affiliation{Racah Institute of Physics, The Hebrew University of Jerusalem, Jerusalem 91904, Israel.}

\author{Jonathan Ruhman}
\affiliation{Department of Physics, Bar-Ilan University, 52900, Ramat Gan, Israel.}

\author{Amit Kanigel}
\email{amitk@physics.technion.ac.il}
\affiliation{Physics Department, Technion-Israel Institute of Technology, Haifa 32000, Israel.}

\begin{abstract}
Finding unambiguous evidence of non-trivial pairing states is one of the greatest experimental challenges in the field of unconventional superconductivity. Such evidence requires phase-sensitive probes susceptible to the internal structure of the order parameter.
We measure the Little-Parks effect to provide clear evidence of an unconventional superconducting order parameter in 4Hb-TaS$_2$.
Namely, we find a $\pi$-shift in the transition-temperature oscillations of rings made of a single crystal. 
We argue that such an effect can only occur if the underlying order parameter belongs to a two-dimensional representation, in other words there are two degenerate order parameters right at the transition.
Additionally, we show that $T_{\rm c}$ is enhanced as a function of the out-of-plane field when a constant in-plane field is applied. Such an increase is consistent with a chiral state, which again, in general only emerges from a two-component order parameter. In combination with previous experiments, our results strongly indicate that \TSB indeed realizes a chiral superconductor.
\end{abstract}

\maketitle
The prospect of topological superconductivity showing exotic quantum phenomena such as protected edge states or fractional vortex states with non-abelian statistics has invigorated the field of unconventional superconductivity~\cite{AndoTSCreview,alicea2012new,alicea2016topological}. 
In order to possess non-trivial bulk topology, the superconductor must be formed out of Cooper-pairs with non-s-wave symmetry.
Unfortunately, most superconducting materials  favor the mundane $s$-wave pairing state. 
%The reason comes from the microscopics. Namely, to get a topologically non-trivial ground state the Cooper pair wave-function itself must be non-s-wave, while  naturally the lowest energy binding  state is featureless. 
While unconventional superconductivity is thus usually associated with correlated electron systems, where strong interactions restrict the pairing channels, understanding the minimal necessary conditions to overcome the natural tendency for conventional superconductivity has proven to be a very difficult task. 
One of the main challenges is to identify materials that exhibit clear experimental evidence of unconventional superconductivity, where theory and experiment can be carefully compared.

The ability to stack atomically thin materials with different properties and at arbitrary relative angles has revolutionized quantum condensed matter research in the last few years~\cite{2D_stacking}. Such heterostructures show intriguing interacting phases such as (high)-$T_{\rm c}$ superconductivty\cite{HighTc_BLG}, correlated insulators \cite{CorrelatedIns_BLG}, many-body excitonic states \cite{ExcitonsRapaport}, structural and electronic ferroelectrics \cite{FerroelectricityMBS,deb2022cumulative}, electronic nematicity~\cite{cao2021nematicity} and magnetism~\cite{chen2020tunable,serlin2020intrinsic}. One of the promising prospects of these heterostructures is the ability to reach new electronic ground states not present in any of the constituent layers. 
An interesting question is whether such heterostructures can be used to manipulate the superconducting pairing channel, resulting in unconventional or even topologically non-trivial superconducting phases.

In this context, the 4Hb polytype of TaS$_2$ provides a particularly interesting example, where  such a ``heterostructure" is naturally occurring. This polytype constitutes a periodic stack of a Mott insulator and candidate spin liquid \cite{ribak2017gapless} (1T-TaS$_2$), and an Ising superconductor (1H-TaS$_2$) \cite{de2018tuning}, which is believed to have an s-wave order parameter~\cite{lian2022intrinsic}.  
The resulting material forms a highly anisotropic  superconductor with $T_{\rm c} = 2.7$ K. The superconducting state exhibits several unconventional properties, including an enhancement of the muon-spin-relaxation ($\mu$SR) rate  below $T_{\rm c}$~\cite{Ribak2020}, a residual T-linear specific heat at low temperatures, zero-energy edge states near step edges observed in STM \cite{Nayak2021evidence} and a mysterious magnetic memory above $T_{\rm c}$, which manifests itself only as spontaneous vortices observed in the superconducting state~\cite{persky2022magnetic}. 

The above unconventional properties have led some of us to argue that the order parameter in \TSB is chiral and hence, topologically non-trivial. Such a state can emerge for order parameters that are degenerate at $T_{\rm c}$ and can indeed explain the $\mu$SR data, the existence of edge states and the spontaneous vortices. However, none of the experiments done so far is capable of determining the order parameter and especially its degeneracy at $T_{\rm c}$.

In this paper, we provide strong evidence of a two-component order parameter in \TSB based on the Little-Parks effect~\cite{Little1962observation} in superconducting rings made of single-crystal \TSB.
First, we find that in roughly half of the rings, a spontaneous $\pi$-junction is formed, manifested as a $\pi$ shift of the oscillation pattern. Such a shift is a direct evidence for an order parameter that changes its sign along the Fermi surface, in other words an unconventional, non-$s$-wave order parameter. Such an effect has previously only been observed in polycrystalline samples~\cite{Tsuei1994pairing, Li2019observation} and requires a two-component order parameter in a single-crystal sample. Second, we find an increase in the critical temperature for out-of-plane fields when a small in-plane field is applied, which can be explained by coupling of the out-of-plane field to a chiral order parameter.

The Little-Parks effect is a  manifestation of the fluxoid quantization in non-simply-connected superconductors.Specifically, $$\Phi'=\Phi+\frac{mc}{n_se^2}\oint_{\mathcal{C}}\vec{j_s}\cdot\vec{dl}=n\Phi_0, $$
where $\Phi$ is the magnetic flux penetrating some contour $\mathcal{C}$ in the ring, $n_s$ is the superfluid density, $\vec{j_s}$ the supercurrent density and $\Phi_0=hc/(2e)$ is the flux quantum. The effect is a consequence of the macroscopic nature of the superconducting condensate, which requires the order parameter to be a single-valued function. Therefore, when the applied flux is not an integer in units of $\Phi_0$ the %superconductor is frustrated
excess flux is screened by a circulating supercurrent, which reduces $T_{\rm c}$. However, each time $\Phi/\Phi_0\in \mathds{Z}$ the full value of $T_{\rm c}$ is restored and as a result, $T_{\rm c}$ oscillates as a function of the magnetic field. While for a conventional superconductor $T_{\rm c}$ has a maximal at zero flux, time-reversal symmetry also allows for a minimum, such that the pattern is shifted by $\pi$. However, such a `$\pi$-ring' requires a sign-changing order parameter~\cite{Geshkenbein1987vortices}.

\begin{figure}[t]
\centering
	\includegraphics[width=1\linewidth]{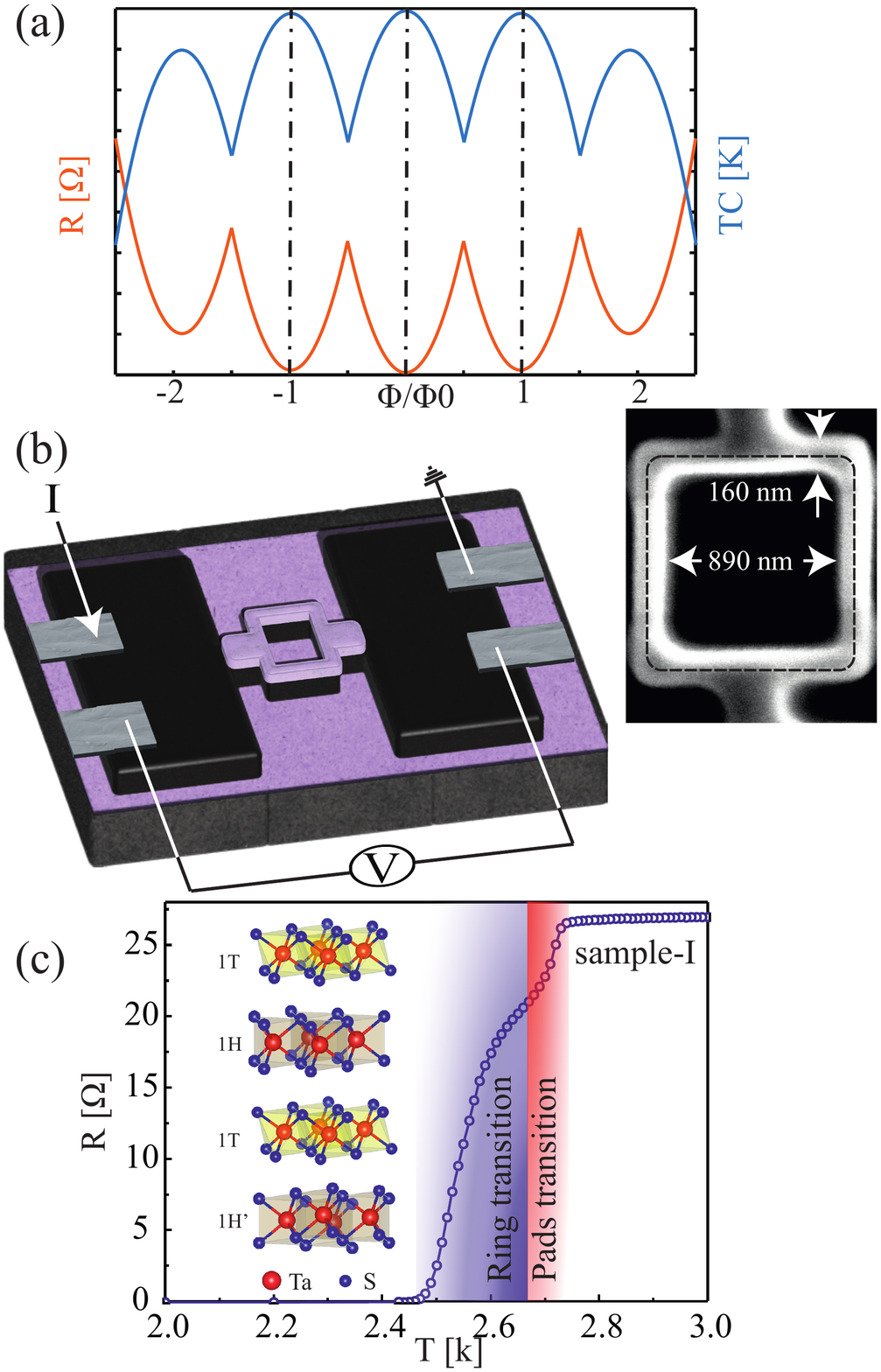}
	\caption{\textbf{The Little-Parks experiment setup} (a) The expected variation of the resistance and of the transition temperature as the flux through the ring is changed.  (b) A schematic description of the device. The purple layer represents the SiO$_x$ layer of the substrate and the protective layer. The aluminum contacts are shown in grey, and the black layer is the 4Hb-TaS$_2$ flake. A scanning electron microscope image of a ring is shown. (c) The temperature dependence of the resistance of sample-I, a 1.1 $\times$ 1.1~$\mu$m$^2$ ring with 140~nm thickness.  The colored regimes represent the transitions of the large pads and of the ring, which take place at slightly different temperatures. Inset: The unit cell of \TSB } 
	\label{Fig1}
	\end{figure}

In practice, it is easier to measure the variation in the resistance of a sample at constant temperature instead of measuring the actual change in $T_{\rm c}$. For samples that show a sharp superconducting transition, the temperature is stabilized within the transition range. Then, small variations in $T_{\rm c}$ as a function of the magnetic field lead to a significant change in the resistance. Fig \ref{Fig1}(a)  shows a cartoon describing the expected variation in $T_{\rm c}$ as a function of the magnetic flux through the sample and the corresponding variation of the resistance.

%% Stopped here

We fabricated several ring-shaped samples made of \TSB single crystals  with sizes ranging between $1.2\times1.2~\mu m^2$ to $0.6\times0.6~\mu m^2$ (see Methods section). Consequently, the expected field periods of the Little-Parks oscillations in the various rings are $\sim$14G to $\sim$60~G. The lateral size of all rings is larger than the coherence length of \TSB ($\sim 35$~nm) and of the order of the magnetic penetration depth ($\sim$ 450~nm).  The rings' thickness ranges from $\sim 90$~nm to $\sim$150~nm. 
Figure \ref{Fig1}(b) shows a scheme of the device and a scanning-electron-microscope image of one of the rings. 

Figure~\ref{Fig1}(c) shows the temperature dependence of the resistance around the superconducting transition for sample I. We find two transitions, one at 2.7~K, which we identify with the transition of the large \TSB pads and a second transition at a slightly lower temperature, which corresponds to the ring itself. The width of the latter transition is about 100~mK, similar to the width of the transition measured in a single crystal. This width sets the amplitude of the Little-Parks oscillations. %In thinner flakes ($\leq$90~nm) the transition width is slightly broader($\sim$250mK) and as a result the oscillations' amplitude is smaller.

\begin{figure*}[]
	\centering
	\includegraphics[width=0.8\linewidth]{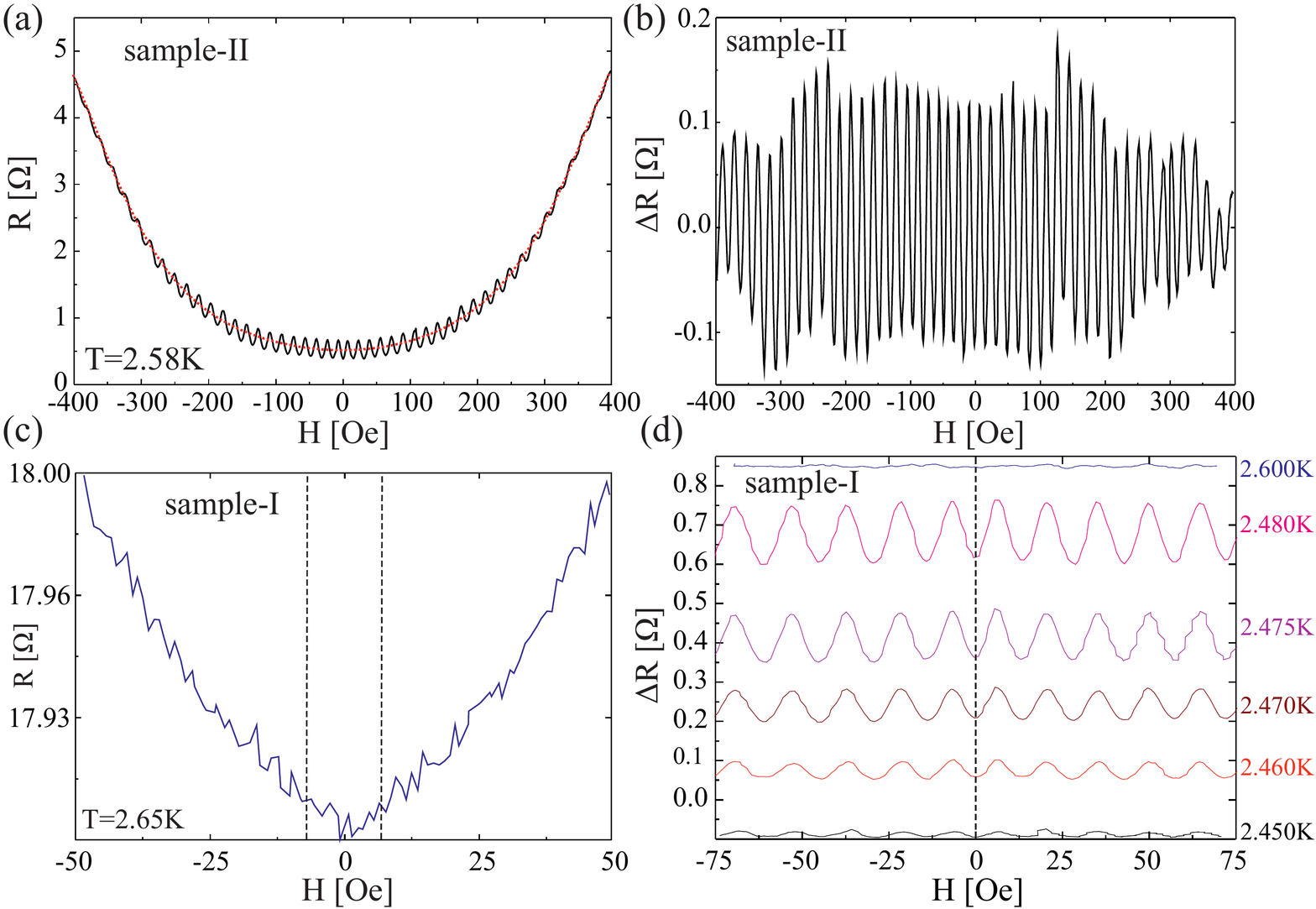}
	\caption{\textbf{ Little-Parks oscillations in \TSB} (a) Little-Parks oscillations for sample-II, a ring with lateral size of $0.9\times0.9~\mu m^2$ and a thickness of 100~nm.  The dashed red line is the forth-order polynomial used for subtracting the background. (b) Same data as in (a) with the background subtracted.  As many as 45 oscillations are measured in this ring. (c) Data for sample-I taken at T=2.65~K (See R vs T for the same sample in Fig. \ref{Fig1} (c)). No oscillations are observed at this temperature. The minimum of the resistance in such curves is used to determine the absolute value of the field in the superconducting magnet. The dashed lines represents the magnetic field at which half of the oscillation period is found in this sample.
	(d) Temperature dependence of the Little-Parks oscillations from  sample-I. The oscillations are observed only in a narrow temperature range.}
	\label{Fig2}
\end{figure*} 

%\red{A main advantage of \TSB\ is the ability to grow pristine single crystals, easily exfoliated with relatively high $T_C$=2.7K\cite{DiSalvo1973preparation,Ribak2020}, and a very narrow transition ($\sim$100mK). Heat capacity and $\mu$-spin-relaxation measurements\cite{Ribak2020} suggest the existence of a fully gaped order parameter, which breaks time reversal symmetry. }

\begin{figure}[]
	\centering
	\includegraphics[trim={0cm 0cm 0cm 0cm},clip,width=0.85\linewidth]{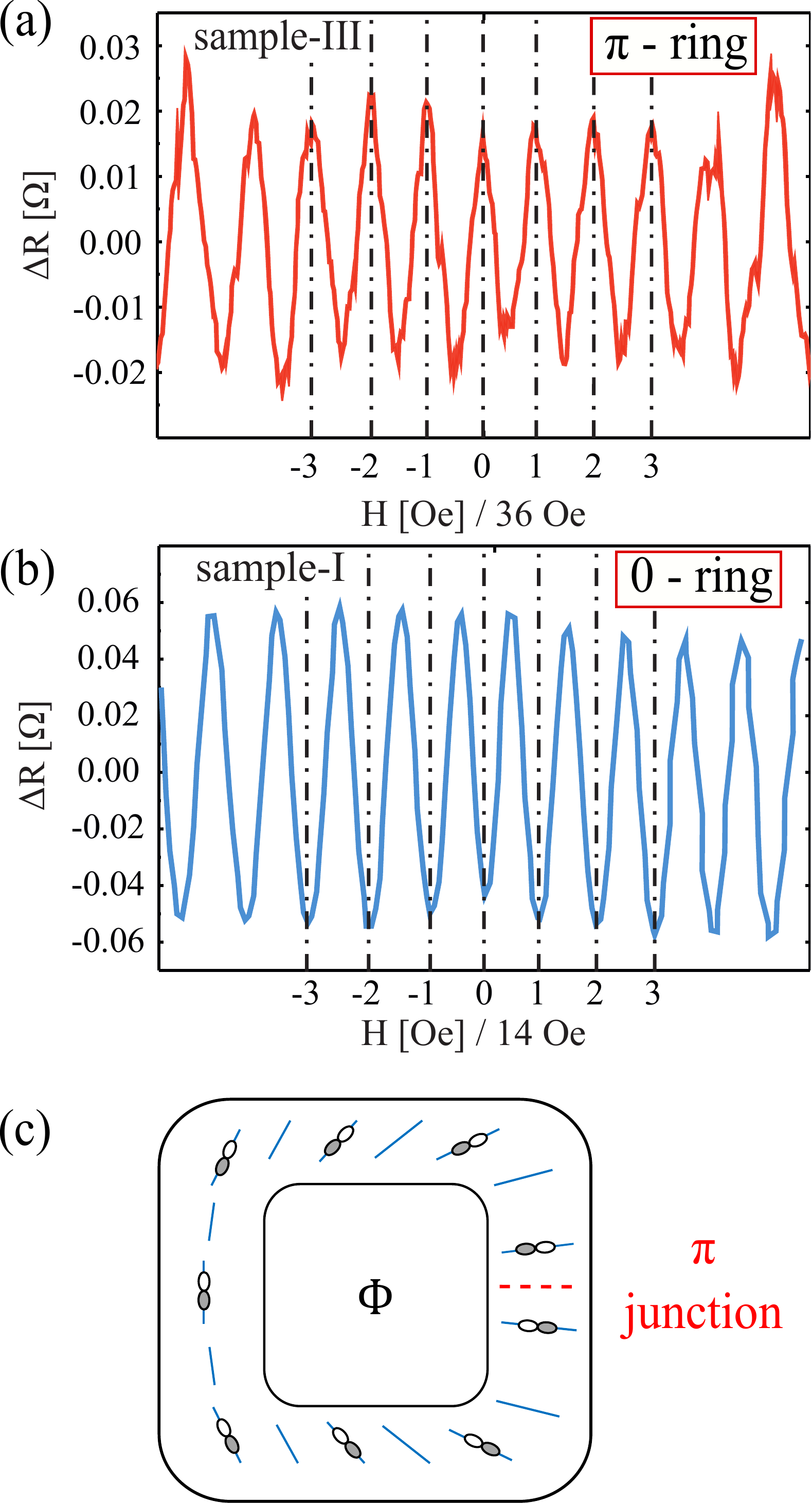}
	\caption{\textbf{$\pi$-shift in the Little-Parks oscillations.} (a) The magneto resistance of a ``$\pi$-ring''. In these rings, the phase of the oscillations  is shifted by $\pi$ showing a  resistance maximum at zero magnetic field. Data shown was measured at T=2.35~K, in a 0.575 $~\mu$m$^2$ ring having a thickness of $\sim$100~nm. (b) The magneto resistance of a  ``0-ring'', having a minimum at zero magnetic field. In both (a) and (b), the background was subtracted. (c) Possible explanation for the $\pi$ shift in the Little-Parks oscillations: Strain fields (blue lines) aligns the two-component order parameter close to $T_{\rm c}$. The strain field presented here realizes a half vortex. Consequently, the order parameter, schematically represented by the white and grey lobes, can not  align with strain without developing a spontaneous $\pi$ junction.  }
	\label{Fig3}
\end{figure}

We find Little-Parks oscillations in all the rings with the expected frequency set by the ring dimensions.
A typical data set is shown in Fig. \ref{Fig2}.
Panel (a) shows the resistance as a function of the field for sample-II, a $0.9 \times 0.9~\mu m^2$ ring, at $T = 2.48$~K. The red line is a 4th order polynomial that is fitted to the data and used for subtracting the background, $R_{\rm bkg} = R_0+R_2 H^2+R_4 H^4$. Panel (b) shows  the Little-Parks oscillations after the background subtraction, $\Delta R = R - R_{\rm bkg}$.
% For small fields, $R_{bkg} = R_0+R_2 H^2$ is sufficient for background subtraction as predicted. The extra $H^4$ term is the result of increased resistivity from destroying the superconducting state in larger fields and therefore must be included in large field measurements.  

% In Fig \ref{Fig2}(c and d) we show the magnetic-field dependence of the magneto-resistance for a $1.1 \times 1.1 \mu m^2$ ring. The R vs T of this ring is shown in Fig.\ref{Fig1}a. 

A crucial step in our experiment is the accurate determination of the zero-field point of the superconducting magnet, which is used to generate the magnetic flux through the ring. To this end, we measure $R(T)$ at a temperature slightly above the bulk $T_{\rm c}$, where the ring is not fully superconducting, but contains superconducting ``islands'' that are susceptible to the magnetic field and give rise to a strong magnetoresistance ~\cite{Tinkham2004introduction}.
In Fig. \ref{Fig2}(c), we show such data for sample-I at $T=2.65$~K, where no oscillations are observed.  The strong magnetoresistance, which is symmetric with respect to zero, allows us to obtain the zero-field point to an accuracy of 1.5~Oe.   

Figure~\ref{Fig2}(d) shows $\Delta R$ in the temperature range in which oscillations are observed. 
%This is a narrow range of about 100mK, the width of the superconducting transition of the ring.
The amplitude of the oscillations we observe is in excellent agreement with theory\cite{moshchalkov1995effect}: For an annular ring with inner radius $R_1$ and outer radius $R_2$, the oscillations in the critical temperature are given by $\mathrm{\frac{\Delta T_{\rm c}}{T_{\rm c}} = \left(\frac{\xi_0^2}{R_1R_2}\right)\left(n-\frac{\pi H R_1 R_2}{\Phi_0}\right)^2}$. 
For this ring, sample-I, we estimate $R_1=550$~nm, $R_2=690$~nm, a coherence length of 35~nm\cite{Nayak2021evidence}, and a critical temperature of $2.7$~K to find the largest change in the critical temperature to be $\Delta T_{\rm c} \approx 2.75$~mK with an oscillation period of 13.5~G. Using the R(T) curve of the ring at 2.47~K, we find oscillations of $\Delta R\approx$140~m$\Omega$.
Thus, the period and amplitude are in excellent agreement with the measured values.

%\red{In this study we use Niobium, a conventional BCS superconductor, and \TSB. For \TSB, we predict within the accepted paradigm\cite{Geshkenbein1987vortices} to find in part of the rings a shift of $\pi$ of the oscillations. The expected behavior is depicted in fig.~\ref{Fig1}(c,d) For Niobium and \TSB\ respectively. }

Figure~\ref{Fig3} presents the main result of this work. Unlike in the Little-Parks experiments in elemental superconductors, we find two different types of rings, 0-rings and $\pi$-rings. For the 0-ring shown in Fig. \ref{Fig3}(b), we find a minimum of the resistance at zero-field as expected. However, the $\pi$-rings show a maximum of the resistance at zero-field. This finding provides a clear signature  of a half-flux-quantum vortex that is spontaneously  created in the $\pi$- rings.

Four out of nine rings we fabricated show a $\pi$ shift. All rings show either a maximum or minimum at zero field, in other words we never observe a fraction of a $\pi$-shift.
The behavior of the different rings is reproducible, meaning that a $\pi$-ring will always show a resistance maximum at zero field, even for consecutive cool downs. We have observed the $\pi$-shift in sample-III even after heating to $365$~K, well above the CDW transition temperature. This disqualifies the electronic charge ordering as a possible origin and strongly suggests that crystal-structure effects play a crucial role in pinning the half-flux vortex.      

%Geshkenbein, Larkin and Barone \cite{Geshkenbein1987vortices} analyzed the conditions for stabilizing  a half-flux vortex in a superconducting ring~\cite{sigrist1992paramagnetic}.
A half-flux vortex in a superconducting ring can be stabilized in different ways, including by combining superconductors with different order-paramter symmetry or polycrystalline rings\cite{Geshkenbein1987vortices,sigrist1992paramagnetic}.
In all cases, the involvement of a non-trivial (non $s$-wave) order parameter is a necessary condition to observe this phenomenon.
For polycrystalline samples, it is enough to have an order parameter that changes sign under rotation and at least three grains. The half-flux vortex forms when the crystal axes of the grains are rotated with respect to each other in such a way that an odd number of Josephson couplings across the grain boundaries are positive. This results in a frustration of the phase locking between the grains, which is relieved by half integer flux. 
A famous example of this effect was measured using a scanning SQUID in the cuprates\cite{Tsuei1994pairing,tsuei1996pairing,tsuei2000pairing}:
The sample was a ring patterned in a YBCO thin film that was grown on a substrate made of three crystals. These crystals were rotated with respect to each other such that the lobes of the $d$-wave order parameter have a negative overlap across one of the three boundaries.  

The rings in our study, however, are cut from a {\it single crystal} and we do not find any evidence for grain boundaries.
The ring fabrication process certainly creates structural imperfections, evident by the fact that the $\pi$-shift is stable in $\pi$-rings. Nevertheless, it is very unlikely that a structural tri-junction that hosts a half-flux vortex was spontaneously created in roughly 50\% of the rings. 
The observation of the half-flux vortex in a single crystal greatly restricts the possible order parameters. We argue in the following that the frustration must come from an internal rotation of the order parameter, which is only possible in a multi-component order parameter. 

\TSB crystallizes in the $\mathrm{P6_3/mmc}$ hexagonal space group (\#194). We thus classify the pairing states according to the point group $D_{6h}$ to determine the coupling to external perturbations such as magnetic field or strain. As \TSB is highly two-dimensional, we restrict ourselves to in-plane pairing, such that we only have to consider four irreducible representations (irreps), namely the two one-dimensional irreps $A_{1g}$ and $B_{1u}$ corresponding to $s$- and $f$-wave pairing, respectively, and the two-dimensional irreps $E_{2g}$ and $E_{1u}$~\cite{fischer:2015a}. For the latter two,
%on the global symmetry of the crystal contains 6 different functions (corresponding to  nearest neighbour pairing states in real space)~\cite{goryo2012possible}. 
%Assuming that the 1T layers that separate the 1H layers reduces significantly the out of-plane hopping, its reasonable to rule out inter-layer pairing. 
%We have 4 possible pairing states: A$_{1g}$, A$_{2u}$, E$_g$ and E$_u$. The last two are two-components paring states. 
the relative phase and amplitude between their two components control the time-reversal- and rotational-symmetry breaking of the order parameter.  
%For example, let us consider the two components of one of the $E$ representations~\footnote{The analysis follows for both representations, or the mixture of. } in their real basis, which we can 
If we denote by $\delta_x$ and $\delta_y$ the real basis of a two-dimensional irrep $E_{1u}$~\footnote{These basis functions indeed transform as $x$ and $y$. Note also that the same arguments hold for the case of $E_{2g}$.}, a general gap function has the form $\hat \Delta(\theta,\phi)= \Delta_0 (\cos \theta\, \delta_x + e^{i\phi}\sin \theta\, \delta_y)\hat{\sigma}^z(i\hat{\sigma}^y)$, where $\hat{\sigma}^i$ are Pauli matrices. Importantly, the spin direction of this spin-triplet gap function is fixed by the strong spin-orbit coupling~\cite{de2018tuning}.
%A general gap can have arbitrary $\theta$ and $\phi$. 
For $\theta = \pi/4$ and  $\phi = \pm\pi/2$, the order parameter is purely chiral and breaks time-reversal symmetry, while a purely ``nematic'' state, which breaks rotation symmetries, is formed by any $\theta$ and $\phi = 0,\pi$.

%This is how far I (Mark) got. I will continue tomorrow.

The existence of $\pi$-rings readily rules out a conventional $s$-wave order parameter ($A_{1g}$). The $f$-wave order parameter ($B_{2u}$) is also highly unlikely in our rings given that they are cut from a single crystal. 
For one of the two-dimensional irreps ($E_{1u}$) and ($E_{2g}$)  the $\pi$-flux effect can be generated by (at least) three domains of the angle $\theta$. The question remains, however, what could cause such domains to appear. The experimental observation that $\pi$-rings remain such in successive cool-downs, even when cycling above the CDW temperature, suggests that the origin is structural. Indeed, the two-component order parameter can couple to crystal strain via the free-energy density term~\cite{sigrist:1991, cho2020,YuanBergKivelson2021}
$ f_{\rm strain} = -\kappa \mathrm{Tr}[ \hat Q \hat\ve] $,
where $\hat \ve$ is the in-plane strain tensor,
$$
\hat Q =|\Delta_0|^2 \begin{pmatrix}
\cos2\theta & \cos \phi \sin 2\theta \\
\cos \phi \sin2\theta & -\cos 2\theta
\end{pmatrix}\,,
$$
and $\kappa$ a coupling constant.
This term is minimized by a real ``nematic'' gap function characterized by $\phi = 0,\pi$ and $\theta$ aligned with the `axis' of the strain. 

Internal strain of the sample can, thus, align the order parameter. As strain in a single-crystal is typically smooth and does not form sharp domain walls, we would naively also expect the order parameter to always be smoothly connected around the ring and as such frustration free. 
However, the strain only defines an axis and not a direction, as apposed to a two-component superconducting order parameter.
Therefore, it is possible to have a strain field that is smoothly connected but not compatible with a smoothly connected superconducting order parameter.
An example occurs when the strain has the topology of a nematic disclination  (see Fig.\ref{Fig3}c). As we show in the supplement, a similar strain field configuration emerges in the vicinity of a crystallographic dislocation.
This scenario is consistent with all our observations, namely (i) the $\pi$-shift remains in successive cool downs, because dislocations are stable topological defects and (ii) we only obersve $0$- or $\pi$-rings, but never any fraction of $\pi$, despite the obervation of time-reversal-symmetry breaking at low temperatures~\cite{Ribak2020}. Importantly, right at the transition, where the Little-Parks experiment is performed, the two components are degenerate and this degeneracy is lifted by strain. Only at lower temperature, the order parameter realizes a chiral combination (see Methods).

% 

% Chiral SC have a phase that changes continuously along the Fermi-surface and can trap a vortex having {\it any} fraction of the flux quanta. This is in contrast to our observation of rings trapping either zero or exactly half of the flux quanta. 

% In nematic SC  the combination of the two component of the order parameter allows a rotation of the nodes direction even when the crystal axes are uniform, this can result in a spontaneous emergence of a $\pi$-junction. 
% The rotation of the axes of the order parameter can be a result of local strain fields. 

% \red{Jonathan, maybe this is the place for your strain half-vortex?}  

%The phase accumulation is not deterministic, and may occur depending on fine-details of the ring\cite{Tsuei1994pairing,Li2019observation}. For odd number of domain walls exist along the ring, an overall phase accumulation of $\pi$ is energetically favorable. For even number of domain walls the "0-ring" state is achieved\cite{Geshkenbein1987vortices,Li2019observation}. %

\begin{figure}[]
	\centering
	\includegraphics[width=1\linewidth]{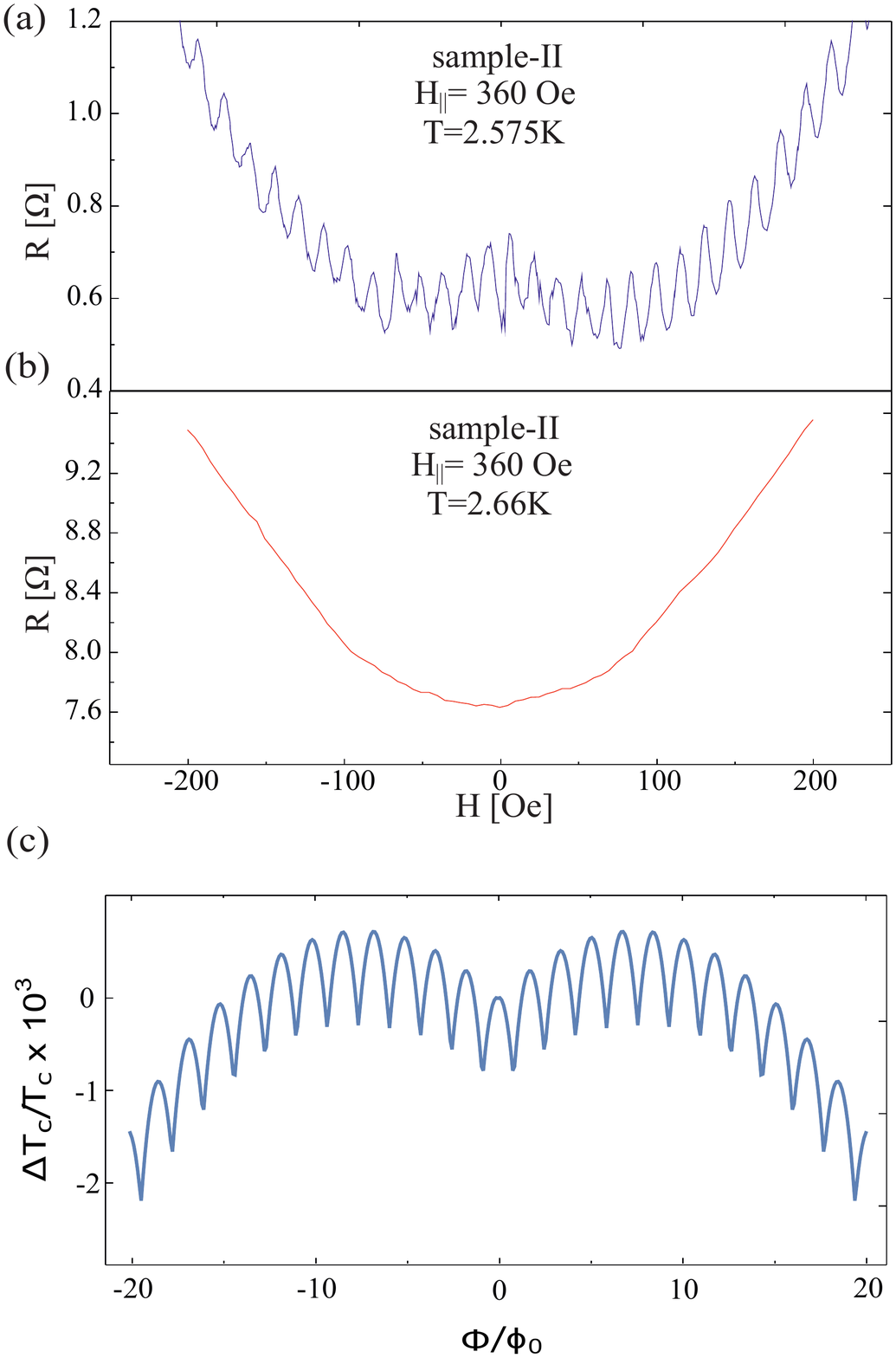}
	\caption{\textbf{Little-Parks oscillations in the presence of an in-plane magnetic field.}  (a) Same ring as in Fig.~\ref{Fig2} (a), measured in the presence of a 360~Oe in-plane magnetic field. The  background has a ``Mexican-hat'' shape in this case with minima at about $\pm$80~Oe.  (b) Resistance as a function of the out-of-plane field with the 360~Oe in-plane applied, at a higher temperature of 2.66~K where the oscillations are absent. The  parabolic background found in the absence of the in-plane field, as shown in Fig. \ref{Fig2}c, is recovered. (c) Theoretical calculation of $\Delta T_{\rm c} / T_{\rm c}$ for an annular ring with similar dimensions and a two-component order parameter. The magnetic field couples to the two-component order parameter linearly and causes the emergence of two maxima. For more details see supplement.  }
	\label{Fig4}
\end{figure}

For chiral superconductors, it was suggested that in-plane magnetic fields can stabilize the formation of half-flux vortices~\cite{Jang2011observation,Roberts2013numerical,Yasui2017little}. 
We thus repeated our Little-Parks measurements in the presence of an in-plane field to check whether we can stabilize more $\pi$-rings. For this purpose, a constant in-plane field was applied and we again measured the resistance as a function of the {\it out-of-plane} field. The results for sample-II are shown in Fig.~\ref{Fig4}. 

%These results raise the question of what would happen in a chiral superconductor? 
%An anticipated result\cite{Vakaryuk2009spin} of a chiral order parameter\cite{Kallin2016chiral} is the existence of a triplet cooper-pairing. This unique pairing may result\cite{Vakaryuk2009spin,Kallin2016chiral} in a half-vortex state in which the critical temperature oscillates with a frequency two times larger than that expected by the ring size\cite{Jang2011observation,Kallin2016chiral,Yasui2017little,Cai2013unconventional}. To achieve this state we subjected the ring to an in-plane magnetic field\cite{Jang2011observation,Cai2013unconventional} using a home-built Helmholtz coils  inserted into the DynaCool system, as shown in fig.~\ref{Fig4}(a). 

Figure~\ref{Fig4}(a) shows the resistance as a function of the out-of-plane field in the presence of a fixed 360 Oe in-plane field. While the resistance still shows a minimum at zero field, in other words the ring remains a $0$-ring, the in-plane field clearly modifies the non-oscillatory contribution to the magnetoresistance\cite{Tinkham1964consequences,Tinkham1963effect,groff1968fluxoid}, which now exhibits a resistance reduction at small fields with a clear minimum at about 80 Oe. This suggests that a small out-of-plane field now increases the critical temperature. 
This should be compared with Fig. \ref{Fig2}(a), which shows the oscillations {\it without} the in-plane field for the same sample. 
In this case the minimum of the background is clearly at zero field and no negative magnetoresistance can be observed. Finally, we point out that in some of the rings we observe the $T_c$ enhancement and $\pi$-shift coincidentally.

In Fig. \ref{Fig4}(b), we show the resistance as a function of the out-of-plane field {\it with} the 360 Oe in-plane field but at a slightly higher temperature, where the Little-Parks oscillations are no longer present. 
We find a similar field dependence to the one observed without in-plane field (see Fig. \ref{Fig2}(c)).  This indicates that this unusual negative magnetoresistance effect is directly related to the onset of a superconducting path around the ring and reflects an increase in T$_{\rm c}$ with the magnetic field. 

% The dependence of $T_c$ on the magnetic field for rings, with a width that is not much smaller than the penetration depth was shown to have two contributions \cite{Tinkham1963effect,groff1968fluxoid} , the first is periodic and yields the oscillations and the second is a parabolic decrease in $T_c$ as the field is increased (see supplementary data). This is the background we observe in our magneto resistance data. 

% In Fig, \ref{Fig4}b on the other hand, the non-periodic part is negative, i.e. leading to a reduction in the resistivity. Thus, we conclude  that $T_c$ is enhanced by about 1.5 mK in the presence of both out-of and in plane magnetic fields compared with zero field, similar in magnitude to the oscillatory part.   

In general, magnetic fields are known to reduce the transition temperature. Only few exceptions are known, such as in very thin Pb films, the 2D metallic interfaces $\mathrm{LaAlO_3/SrTiO_3}$ \cite{Gardner2011enhancement}, or recently in twisted double bilayer graphene~\cite{liu2020tunable} and bilayer graphene~\cite{zhou2022isospin}. In particular, the coupling of the field to the spins has been suggested to lead to such an effect~\cite{lee2019theory}. Given the strong Ising spin-orbit coupling in this material, such an origin is highly unlikely in this system.
%There it has been proposed that the origin of this effect is a spin-polarized paired state. However, such a pairing state is unlikely to exist in \TSB due to the strong Ising spin-orbit coupling. 
%as the case of superfluid $\mathrm{He^{3}}$. For such superconducting state the order parameter is written as $\Delta_{\uparrow\uparrow} + \Delta_{\downarrow\downarrow}$, and can be thought of as two non-interacting condensates~\cite{Kallin2016chiral}. For such a phase the magnetic naturally increasing the critical temperature~\cite{lee2019theory}. However, the Ising spin-orbit coupling in \TSB makes this scenario highly unlikely.
A more plausible explanation here is the orbital coupling between a chiral order parameter and a magnetic field 
$
f_{\rm chiral} = -K_{\rm c} H |\Delta_0|^2 \sin 2\theta \, \sin \phi
$. Such a term favors a chiral state $\theta = \pi/4$ and $\phi = \pm \pi/2$ enhancing $T_{\rm c}$ linearly with the absolute value of the magnetic field and only appears for two-component order parameters (details see Methods section). 
In Fig.~\ref{Fig4}(c), we plot a theoretically calculated Little-Parks oscillation pattern with such a term, which is in good qualitative agreement with the data. 

{\it Conclusions--} Measuring the Little-Parks effect, we provide convincing evidence that the superconducting order parameter in \TSB belongs to a two-component representation. In particular, we present two main findings, both of which are most easily explained by a two-component order parameter. First, The Little-Parks oscillations of roughly half of our rings exhibit $\pi$-shift in spite of the fact that the rings are formed out of a single crystal. Second, the envelope of the resistance oscillations shows a clear minimum at a finite field (for a constant 360Oe in-plane field), indicating an increase in $T_{\rm c}$ for finite applied fields.

We propose that the $\pi$-shift arises due to strain fields that couple to the two-component order parameter and cause it to form half a rotation around the ring. Elastic theory shows that similar strain fields emerge around a dislocation. The origin of such dislocation defects is not currently known. It may result from the sample growth procedure, which essentially includes a rapid cool down, or the stresses applied on the ring by the pads and substrate due to thermal compression in the cooling (see methods). It is also not clear why the defects occur in roughly half of the devices. These questions are left to future work. 

Finally, we have argued that the increase in $T_{\rm c}$ as a function of perpendicular magnetic field indicates a chiral component, which is induced by the field itself. 
However, the fact that this $T_{\rm c}$ enhancement  is only seen with an in-plane field is not naturally explained by this scenario. 
In the Methods section, we show that when both strain and field are present the order parameter is a mixture of chiral and nematic states with $\phi = \pi/2$ and $0<\theta<\pi/4$. In this scenario, the  enhancement of $T_{\rm c}$ with field is suppressed due to the competition with  strain.  Thus, we may speculate that if the coupling to strain is suppressed by the in-plane field, it will lead to the observed $T_{\rm c}$ enhancement. 

The consistent picture thus emerging from our experiments is the following: Right at $T_{\rm c}$, \TSB enters a superconducting phase that belongs to a two-dimensional irreducible representation. Close to $T_{\rm c}$ this degeneracy is lifted  by internal strain in the sample, which can lead to both $0$- and $\pi$-rings. Only upon lowering the temperature further,
the system breaks time-reversal symmetry spontaneously by forming a chiral order parameter. We note that recent measurements of an anisotropic in plane $H_{c2}$ also support this scenario, where it was argued that the order parameter is a mixture of chiral and nematic state tuned by temperature~\cite{Yoram2022}. The microscopic origin of the two-component order parameter thus remains an outstanding open question.  

{\it Acknowledgements--} We thank Jorn Venderbos, Patrick Lee, Avraham Klein, Daniel Agterberg and Erez Berg for very helpful discussions. 
We thank Itay Mangel, Noa Somech and Shay Hacohen-Gourgy for help with the experiment. 
 JR was supported by the Israeli Science Foundation grant no. ISF-994/19.
 AA, IF and AK were supported by the Israeli Science Foundation grant no. ISF-1263/21

 \appendix
 
 \section{Methods}
 \subsection{Sample preparation}
 High-quality single crystals of 4Hb-TaS$_2$ were grown using the chemical vapour transport (CVT) method. A stoichiometric mixture of tantalum (Ta) and sulfur (S) was sealed in a quartz ampoule under vacuum.
  1\% of Se was added to the mixture. It was found to significantly improves the sample quality. The mixture underwent a sintering process, forming a boule of 4Hb-TaS1.99Se0.01. The boule was crushed and placed in a 200-mm-long quartz ampoule with 16-mm diameter. Iodine was added as a transport agent, and the ampoule was sealed under vacuum. The ampoule was then placed in a three-zone furnace, where the hot ends were heated to 800$^\circ$C and the middle part was kept at 750$^\circ$C. After about 30 days, the ampoule was quenched in cold water.
   Crystals having a few mm in size are found in the cold part of the ampoule.

The crystal structure and chemical composition were verified using X-ray diffraction and electron energy dispersive spectroscopy in a scanning electron microscope. The measurements show that  the actual amount of Se in the crystals is lower than 1\%

 \subsection{Rings fabrication}
 For this experiment we fabricated \TSB rings with typical dimensions of 1$\mu$m. We exfoliate \TSB flakes on a $\mathrm{SiO_2}/\mathrm{Si}$ substrate using the standard dry transfer   technique. Ti/Al contacts are evaporated using electron beam lithography. We cover the flakes with a protective layer of $\mathrm{SiO_2}$ $in-situ$ before using a FEI Helios NanoLab DualBeam G3 UC focus ion beam (FIB) to carve the desired ring from the exfoliated flake. 
 \subsection{Magneto-resistance measurements}
 We measure the rings resistance by averaging 200 IV curves for each magnetic field, with a maximal current of 250~nA using a Quantum-Design DynaCool system with temperature stability of $\pm180\mu$K. Extending the IV curve measurements to higher currents did not change the results, in agreement with measurement in different systems \cite{Li2019observation}. 
 
  \subsection{Finding the zero-field in the superconducting magnet}
  
   \begin{figure}
	\centering
	\includegraphics[width=0.8\linewidth]{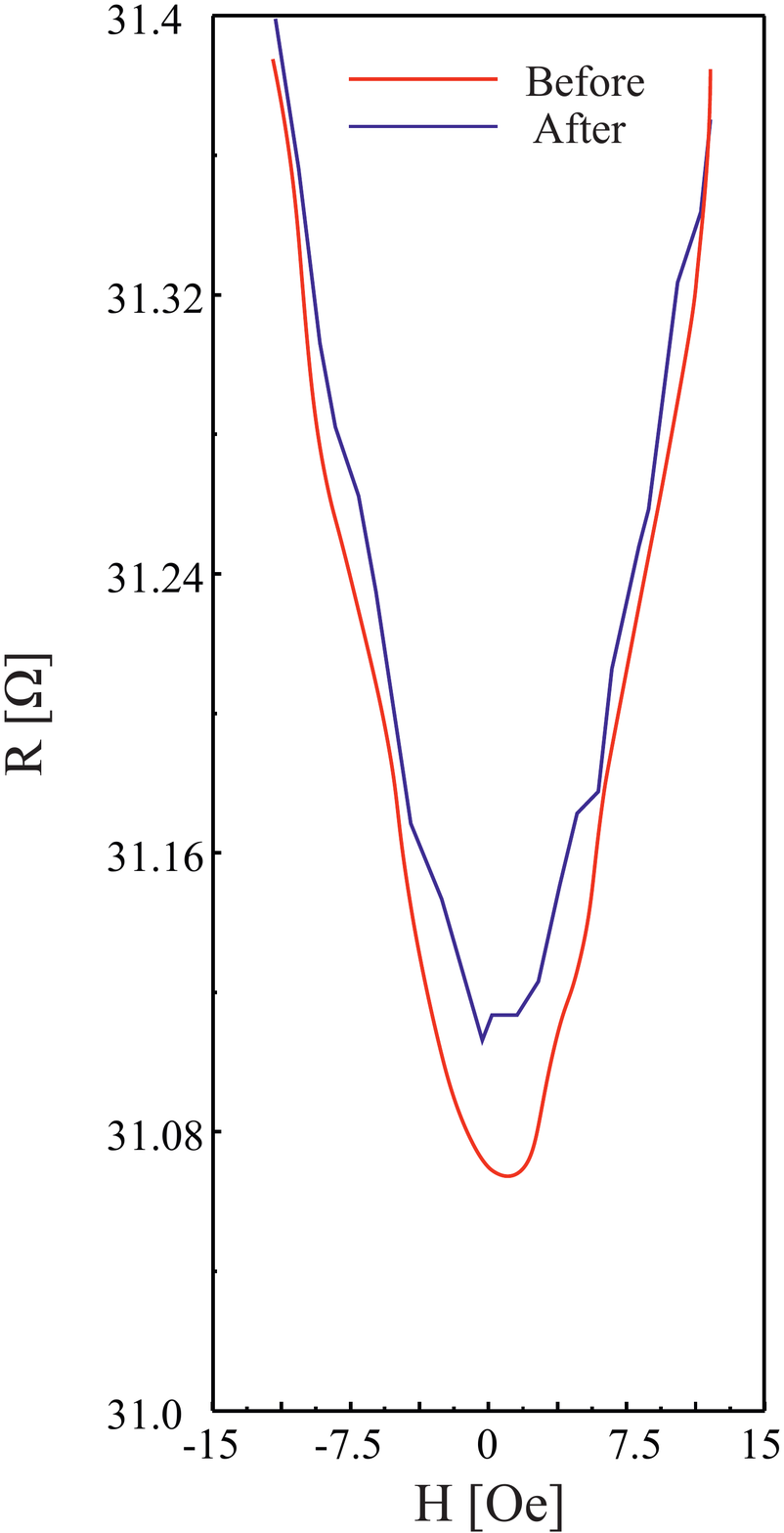}
	\caption{\textbf{Finding the zero-field.} Resistance as a function of the magnetic field for sample-III. The two measurements were done before and after the oscillations shown in Fig. \ref{Fig3}(a) were measured. }
	\label{BKG}
\end{figure} 
 
  We use a superconducting magnet, trapped flux makes it difficult to find the "real" zero-field. 
  
 Before each experiment we minimize the trapped flux by oscillating the magnetic field to +1T, -0.5T, +1kG, -500G , +100G and finally 0.
  
 For each sample we first stabilize the  temperature within the superconducting transition so that it shows  a strong magneto-resistance  but no oscillations, see figure~\ref{Fig2}(c). We then measure R(B) for a range of magnetic field around "zero". The  zero-filed is set to be the field at which the resistance is minimal.
  
  We found that as long as the magnetic fields are smaller than 400G, the residual field does not change and remains smaller than 5G.
  
  We measure the residual magnetic field in the system before and after each Little-Parks measurement. In  figure~\ref{BKG} we show the resistance as function of the field measured at T=2.7K for sample-III. We show the curves measured before and after the temperature was lowered and oscillations were measured. No shift can observed. In this $\pi$-ring the oscillation period is 36Oe, clearly the uncertainty in the  field is much smaller than half of the oscillation period.  
  
  We emphasize that our uncertainty in the size of the magnetic field  is smaller than half of the Little-Parks oscillation period for all the  samples that we measured.   
  
   \section{Theory}
    In the following, we theoretically discuss the experimental observation of the $\pi$-flux and the enhancement of $T_{\rm c}$ in  a field based on the assumption that the superconducting order parameter belongs to a two-component representation of the symmetry. We also discuss how the observed signatures cannot be explained with a single-component order parameter without fine tuning and/or unrealistic assumptions.
    
  \subsection{Ginzburg-Landau Theory of the two-component order parameter}
 
Assuming a two-component order parameter of the form
\begin{align}
    \hat \Delta_{\bs k}(\theta,\phi)= \Delta_0 \left[\cos \theta\, \delta_x(\bs k) + e^{i\phi}\sin \theta\, \delta_y(\bs k)\right](-i\hat{\sigma}^y)\hat{\sigma}^z \,,
\end{align}
which is written in the space of $\Psi_{\bs k}^\dag = (\psi_{\bs k\uparrow}^\dag,\psi_{\bs k\downarrow}^\dag)$, such that the order parameter is given by $\langle \Psi_{\bs k}^T \hat \Delta (\bs k)\Psi_{-\bs k}\rangle $. Here $\delta_{x,y}(\bs k)$ are momentum-dependent basis functions belonging to the two-dimensional irreducible representation $E_{1u}$. Note that these transform like the $x$ or $y$ coordinates, but can have a more complicated form. Also, while we have chosen here the spin-triplet (inversion-odd) order parameter for concreteness, the following discussion also holds for an inversion-even order parameter transforming as $E_{2g}$. The angles $\theta$ and $\phi$ are internal degrees of freedom of the order parameter, which will appear in the corresponding Ginzburg-Landau (GL) theory (not to be confused with the superconducting phase, which is implicitly encapsulated in $\Delta_0 = |\Delta_0|e^{i\varphi}$). The correpsonding Ginzburg-Landau free energy density can then be written in terms of a two-component order parameter $\bs\eta = (\eta_1, \eta_2)$, where the two components relate to the above gap function as $\eta_1 = \cos\theta \Delta_0$ and $\eta_2 = e^{i\phi} \sin\theta \Delta_0$. In the following, we thus start from the free energy density
\begin{widetext}
\begin{align}\label{eq:app GL theory}
    f[\bs \eta] = K_1 \left(|\bs D \eta_1|^2 +|\bs D \eta_2|^2\right)+ K_2 |\bs D\cdot \bs \eta|^2+K_3 |\bs D\times\bs\eta|^2 + \alpha(T - T_c)|\bs \eta|^2 + \beta_1 |\bs \eta|^4 + \beta_2 |\bs\eta^*\times \bs \eta|^2 -\kappa \mathrm{Tr}[ \hat Q \hat\ve]\,,
\end{align}
 where $\bs D = -i \nabla + {e} \bs A$ is the (in-plane component of the) covariant derivative (with $c=\hbar=1$). For simplicity, we have neglected terms that reflect crystal symmetry breaking and assumed a rotationally symmetric model. Finally, the last term, propositional to $\kappa$, describes the coupling to strain, where
$$
\hat Q =\begin{pmatrix}
|\eta_1|^2 - |\eta_2|^2 & \eta_1^*\eta_2 + c.c. \\ 
\eta_1^*\eta_2 + c.c. & |\eta_2|^2 - |\eta_1|^2
\end{pmatrix}\,
$$
 and 
 $$
 \hat \ve = \begin{pmatrix}
\ve_{xx}-\ve_{yy} & \ve_{xy} \\
\ve_{xy} & \ve_{yy}-\ve_{xx}
\end{pmatrix}\,.
 $$

Previous experiments~\cite{ribak2017gapless,Nayak2021evidence,persky2022magnetic} are consistent with a fully gapped and chiral superconducting state. We may assume that $\beta_2<0$. Moreover, we note that for any $K_2\ne -K_3$ there is a linear coupling between the transverse magnetic field and the superconducting order parameter. 

\subsection{The Little-Parks effect in uniform strain}
We first analyze the Little-Parks effect in the presence of uniform strain along $\hat{\bs x}$, namely $\ve_{xx} = -\ve_{yy} = \ve_0/2$ and $\ve_{xy} = 0$. Moreover, we assume the ring to be annular with inner and outer radii given by $R_{1}$ and $R_2$, respectively. Finally, we first neglect coupling of order parameters by the magnetic field, which is achieved by setting $K_2 = -K_3$. We will relax this assumption in the next subsection. 

Close to $T_{\rm c}$, where the quartic terms are negligible, and in the absence of a magnetic field, the order parameter will be aligned by the strain term, such that we have $\eta_1 = \eta_0$ and $\eta_2 = 0$. The resulting free energy density then takes the form 
\begin{align}
f[\bs \eta] = (K_1+K_2)|D_x \eta_1|^2+K_1|D_y \eta_1|^2  +\left[\alpha (T - \tilde T_c) + \kappa \ve_0\right]|\eta_1|^2+ \mathcal O(\eta^4)\,.
\end{align}

 \end{widetext}
 
 We can include a magnetic field via the gauge choice $\bs A = H(-y\hat{\bs x}+x\hat{\bs y})/2 = Hr/2 \hat{\bs \g}$ in cylindrical coordinates with $\g$ the azimuth. We also assume that the gap function does not depend on the radius and has a winding number $n$, such that $\eta_0 = |\eta_0|e^{in\g}$. 
The free energy up to quadratic order is then given by 
 \begin{align}\label{app:eq:free-energy}
     F = \int r dr d\g dz f[\eta_0] = \alpha \pi (R_2^2 - R_1^2)h \left[T-\tilde T_{\rm c}(H)\right]|\eta_0|^2,
 \end{align}
 where $h$ is the height of the ring and
\begin{align}\label{eq:app:Tc_unif}
\tilde T_{\rm c}(H) = T_{\rm c} + {\kappa \ve_{xx}\over \alpha}- g \int_{R_1}^{R_2}{dr\over r} \left( n-{e H r^2 \over 2} \right)^2\,,
\end{align}
with $g = { (2K_1+K_2)/ 2\alpha (R_2^2 - R_1^2)}$.
As usual, $n$ is chosen to minimize $F$ (or maximize $\tilde T_{\rm c}$). 

\begin{figure}[]
	\centering
	\includegraphics[width=0.8\linewidth]{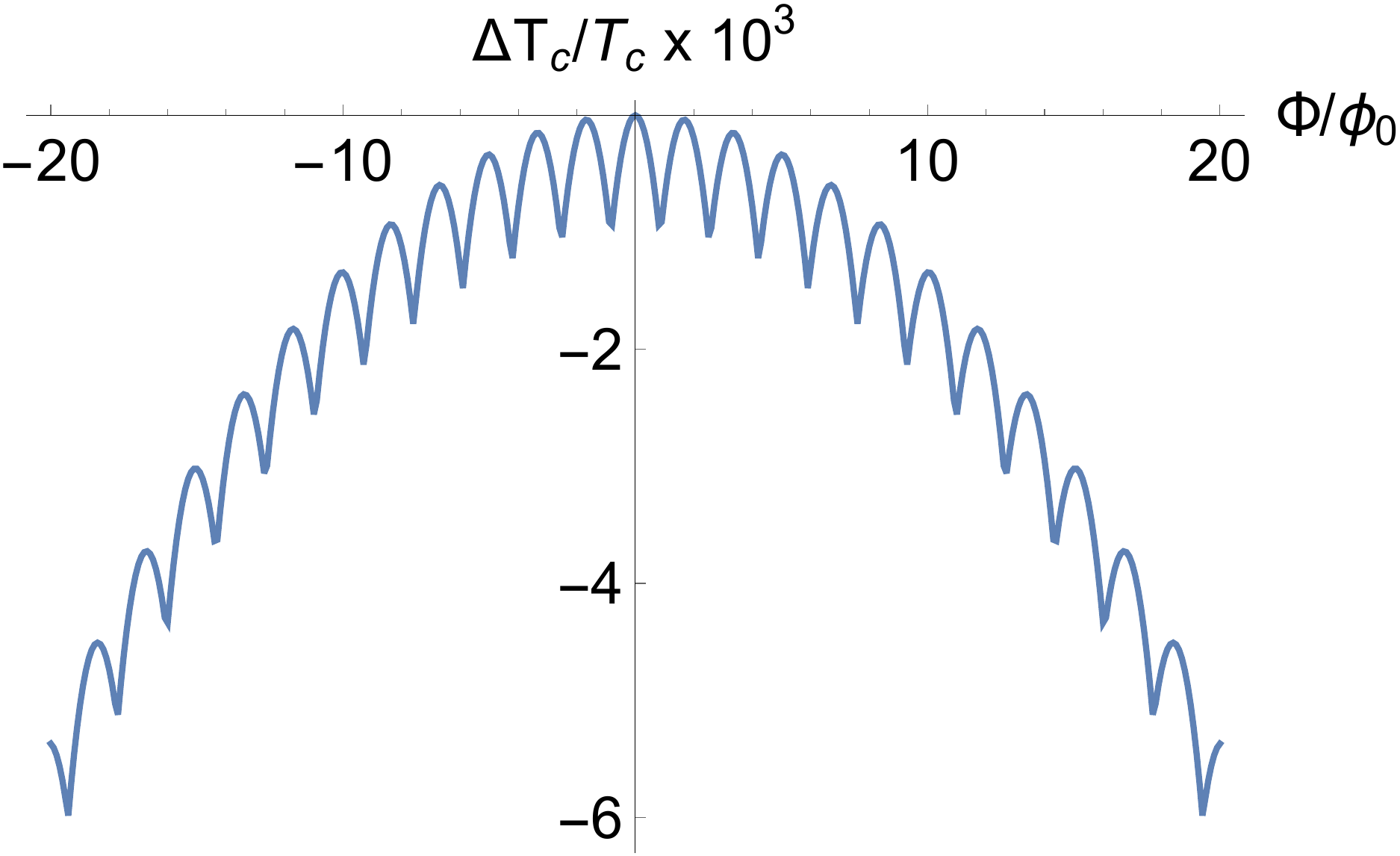}
	\caption{The value of $\tilde T_{\rm c}$ \eqref{eq:app:Tc_unif} vs. magnetic flux $\Phi/\phi_0$. Here we used $g= 0.01 (T_{\rm c}+\kappa \ve_{xx}/\alpha)$ and $R_1 / R_2 = 0.85$. Also note that $\Phi \equiv \pi R_1^2 H$.}
	\label{fig:supp:theory unif strain}
\end{figure}

Figure~\ref{fig:supp:theory unif strain} shows Eq.~\eqref{eq:app:Tc_unif} with flux $\Phi/\phi_0$, where $\Phi \equiv \pi R_1^2 H$. We use the radii ratio $R_1 / R_2 = 0.85$ and set the parameters $g= 0.01 (T_{\rm c}+\kappa \ve_{xx}/\alpha)$. The parameters are chosen to fit the dimensions of the ring in the experiment and the size of the oscillations (order $10^{-3} T_{\rm c}$). 
Interestingly, the strength of the parabolic envelope and the size of the oscillations are not independent of each other. 
As can be seen in the figure, the parabolic envelope of the oscillations fits reasonably well to that seen in Fig.~\ref{Fig2}. Namely,  $\D T_{\rm c}/T_{\rm c}$ reaches about $-10^{-3}$ after an order of 10 oscillations. This shows that the main origin of the parabolic magnetoresistance comes indeed from the finite width of the ring.

As a final note, we mention that Eq.~\eqref{app:eq:free-energy} with $K_2=0$ describes any single-component order parameter, in other words an order parameter transforming like a one-dimensional irreducible representation. This is not surprising, as so far we have described a $0$-flux ring.

\subsection{The Little-Parks effect in a chiral state}
Another simple limit, is the case where there is no strain. Then, the magnetic field will naturally select the chiral state (assuming that $K_2\ne -K_3$). In this case, $\eta_1 = \eta_0$ and $\eta_2 = \pm i\eta_0$, where $\pm$ denotes the chirality. The GL free energy density now assumes the form 
\begin{widetext}
\begin{align}
    f[\eta_0] = (2K_1+K_2+K_3)|\bs D \eta_0|^2- e|H|(K_2+K_3)|\eta_0|^2+\alpha (T-T_{\rm c}) |\eta_0|^2+\mathcal O(\eta^4)\,,
\end{align}
where the absolute value of $H$ comes from choosing the chiral state, which is favored by the magnetic field.

As before, we will assume that $\eta_0 = |\eta_0|e^{in\gamma}$ and $\gamma$ is the azimuth. Integrating over the volume of the ring, we obtain Eq.~\eqref{app:eq:free-energy} with $\tilde T_c(H)$ given by 
\begin{align}\label{eq:app:Tc_chiral}
\tilde T_{\rm c}(H) = T_{\rm c} + {2 \eta_0^2(K_2+K_3)\over \alpha R_1^2}{|\Phi| \over  \phi_0 }- g'\int_{R_1}^{R_2}{dr\over r} \left( n-{e H r^2 \over 2} \right)^2\,,
\end{align}
where $g' = {(2K_1+K_2+K_3) /2\alpha (R_2^2 - R_1^2)}$.
 \end{widetext}

\begin{figure}[]
	\centering
	\includegraphics[width=0.8\linewidth]{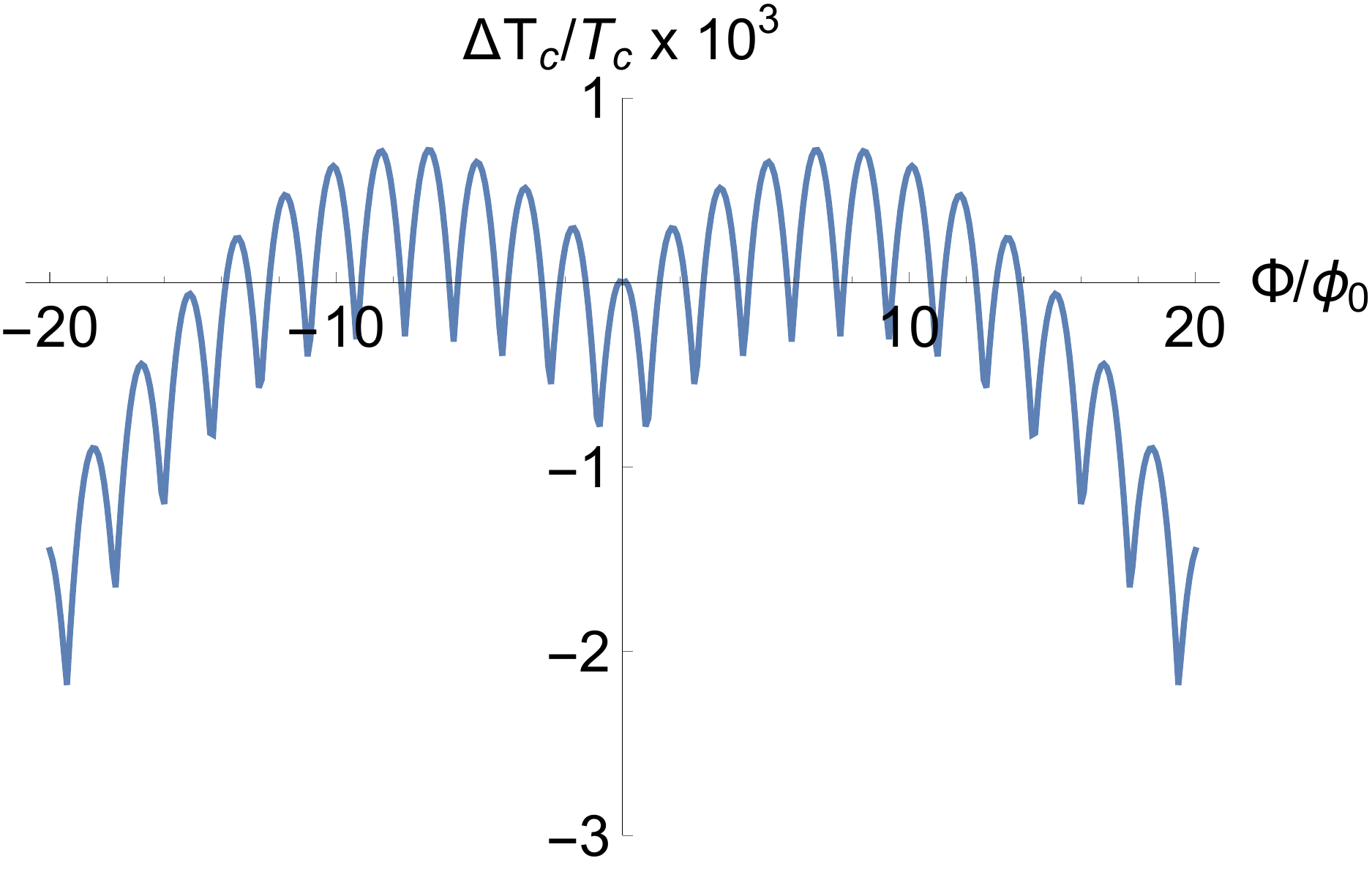}
	\caption{Eq.~\eqref{eq:app:Tc_chiral} vs. $|\Phi|/\phi_0$ for $g' = 0.01 T_c$ and $K_2+K_3  = 2\times 10^{-4} \alpha R_1^2 T_c$. This result should be compared with panel (b) of Fig.~\ref{Fig4}. }
	\label{fig:supp:theory chiral 1}
\end{figure}

\begin{figure}[]
	\centering
	\includegraphics[width=0.8\linewidth]{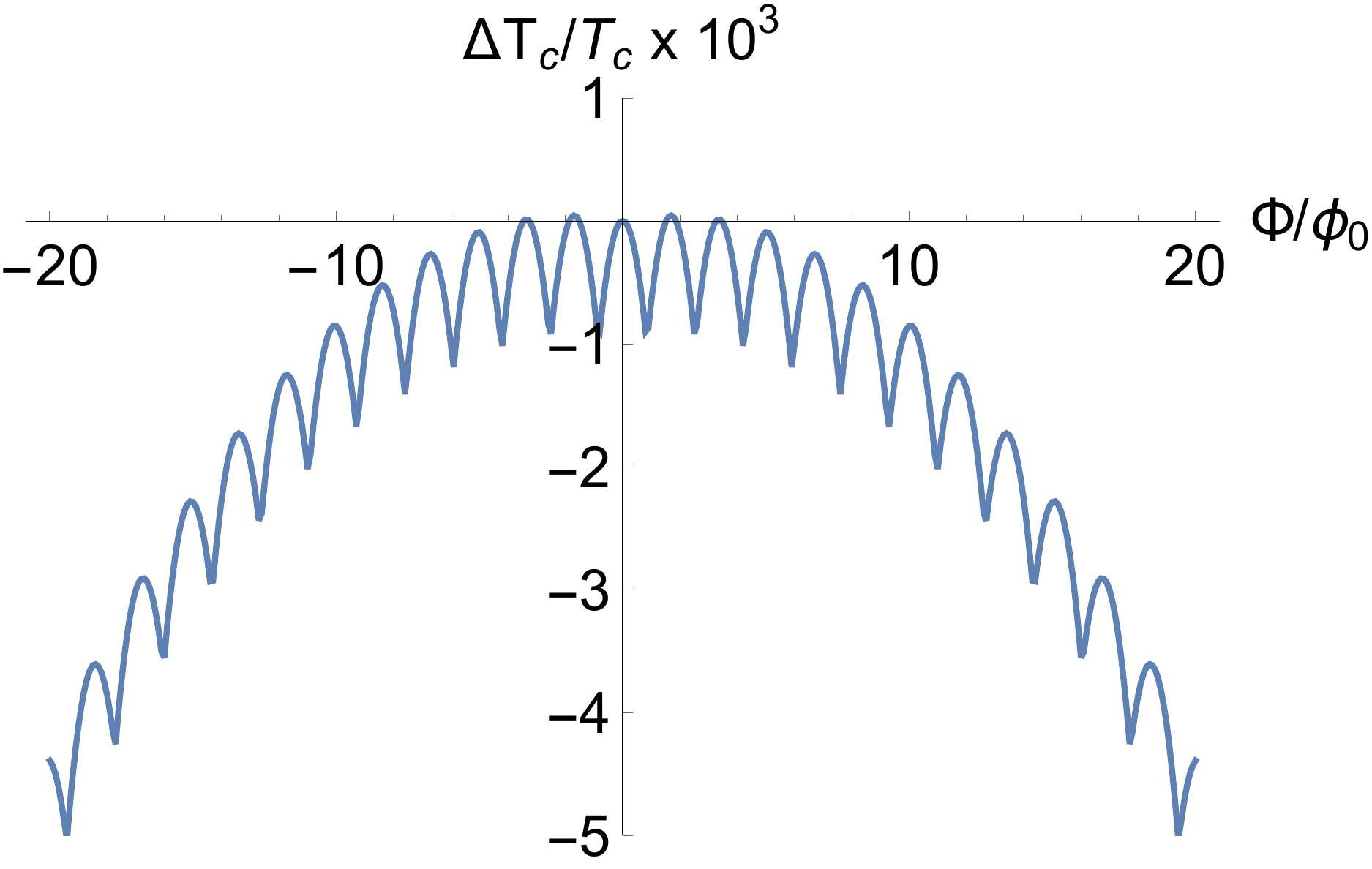}
	\caption{Eq.~\eqref{eq:app:Tc_chiral} vs. $|\Phi|/\phi_0$ for $g' = 0.01 T_c$ and $K_2+K_3  = 0.5\times 10^{-4} \alpha R_1^2 T_c$ }
	\label{fig:supp:theory chiral 2}
\end{figure}

In Fig.~\ref{fig:supp:theory chiral 1} and \ref{fig:supp:theory chiral 2} we plot Eq.~\eqref{eq:app:Tc_chiral} for $g' = 0.01 T_{\rm c}$, $K_2+K_3  = 2\times 10^{-4} \alpha R_1^2 T_{\rm c}$ and $K_2+K_3  = 0.5\times 10^{-4} \alpha R_1^2 T_{\rm c}$, respectively. Fig.~\ref{fig:supp:theory chiral 1} shows an enhancement in $T_{\rm c}$, seen when a magnetic field is applied. This resembles the situation in Fig.~\ref{Fig4}, where an in-plane magnetic field is applied. Fig.~\ref{fig:supp:theory chiral 2} shows the same effect with a weaker value of $K_2+ K_3$. This plot should be compared with panel (a) of Fig.~\ref{Fig2}. Indeed, we find that the parabolic dependence near zero field in Fig.~\ref{Fig2} is flatter than the expected behavior for a non-chiral order parameter Fig.~\ref{fig:supp:theory unif strain}. 

As such, while the difference between the values of the parameter $K_2+K_3$ in the two figures explains the experiment, it is highly unlikely that an in-plane magnetic field causes such a shift. The parameters of the GL free energy are typically set by the non-interacting band structure which is not affected by such a small field. It is much more likely that the in-plane magnetic affects the competition between the strain, which prefers a real order parameter, and the out-of-plane magnetic field, which prefers a chiral state.

 
%%%%%%%%%%%%%%%%%%%%%%%%%%%%%%%%%%%%%%%%%%%%%%%%%%%%%%%%%%%%%%%%%%

\subsection{The chiral-nematic mixed state in the presence of strain and magnetic field} 
In the previous two subsections, we have chosen gap functions that are either purely real with $\phi = 0,\pi$ (``nematic'') or $\phi = \pi/2$ (``chiral'').  We now comment on the more realistic situation of both non-zero strain and $K_2\neq -K_3$. Without an out-of-plane magnetic field, the degeneracy of the order-parameter is broken and the order parameter will be `aligned' with the strain and TRS is preserved as discussed above. However, the coupling to a TRS-breaking order parameter through an out-of-plane magnetic field is still present in the free energy, and as such the system can still have an increased $T_{\rm c}$ at small fields. Specifically,
%In the general case however, where a finite out of plane magnetic field and strain are simultaneously present, the gap will be a mixture of the two. In this case, 
$T_{\rm c}$ is controlled by the quadratic GL free energy density (neglecting spatial variations of the gap)
\begin{align}
    f = \boldsymbol \eta^\dag 
    \begin{pmatrix}
    \alpha(T-T_c) - \kappa \ve_{xx} & i e(K_2+K_3)H \\
    -i e(K_2+K_3)H &  \alpha(T-T_c) + \kappa \ve_{xx}
    \end{pmatrix}
    \boldsymbol \eta + \mathcal{O}(\eta^4)\,,
\end{align}
where we have assumed the strain is uniform and along the $x$ direction. 

$T_{\rm c}$ is given by the largest negative eigenvalue of the matrix, that is
\begin{align}
   \tilde T_c = T_c + {1\over \alpha}\sqrt{\kappa^2 \ve_{xx}^2 + e^2(K_2+K_3)^2H^2 }.
\end{align}
Thus, due to  the competition with strain, the dependence of $T_{\rm c}$ on magnetic field  will be quadratic close to $H = 0$, as apposed to the linear dependence assumed in Eq.~\eqref{eq:app:Tc_chiral}.  The state right below $T_{\rm c}$ is then a ``nematic-chiral'' mixture with $\phi = \pm \pi/2$ (depending on the direction of the field) and $$\cos \theta = {1\over \sqrt{2}}\left[1+ {\kappa \ve_{xx}\over \sqrt{\kappa^2 \ve_{xx}^2+e^2 (K_1+K_2)^2H^2}}\right]^{1/2}\,.$$

This mixed state is expected to have a weaker positive contribution to $T_{\rm c}$ due to the chiral state as compared to Eq.~\eqref{eq:app:Tc_chiral}. Namely, when $e H (K_2 + K_3) \ll \kappa \ve_{xx}$ this contribution will be quadratic in $H$. Given that $T_{\rm c}$ is expected to increase as $H^2$ due to the finite width of the ring, this chiral contribution is only expected to reduce the coefficient in front of $H^2$, which will still be positive, thus not showing an overall $T_{\rm c}$ enhancement. 

Again, we do not know the origin of the $T_{\rm c}$ enhancement when an in-plane field is present. However, we can speculate that if the in plane field suppresses the coupling to strain then this quadratic contribution can be converted to a dominant linear contribution as in Fig.~\ref{eq:app:Tc_chiral} when the in-plane field is present. On the other hand, when it is not present, the strain is dominant thus suppressing the $T_{\rm c}$ enhancement. 

 Finally, note that a coupling of the magnetic field to two order parameters is more generally allowed. In particular, noting that the field transforms as $A_{2g}$, any order parameter combination that transforms as $A_{2g}$ is allowed, since $A_{2g}\otimes A_{2g} = A_{1g}$, in other words the full combination transforms as a scaler. As an example, an order parameter of $B_{1u}$ symmetry can couple to one with $B_{2u}$ symmetry through the magnetic field. For combinations of higher-dimensional irreps, the decomposition needs to contain $A_{2g}$. This is in particular possible for the two-dimensional irreps relevant for \TSB, since $E_{2g} \otimes E_{2g} = A_{1g} \oplus A_{2g} \oplus E_{2g}$ and $E_{1u} \otimes E_{1u} = A_{1g} \oplus A_{2g} \oplus E_{1u}$. As just noted, the effect will only be appreciable, if the two coupled order parameters are very close in energy, in other words have (almost) degenerate $T_{\rm c}$. While this is naturally given for the two-dimensional irreps even with a small perturbation such as strain, an $f$-wave order parameter has no natural partner in this system, and no increase in $T_{\rm c}$ can be expected.

\subsection{The strain induced by a dislocation}
In this section we seek the fundamental solution of stress in a two-dimensional ring geometry in the presence of a dislocation defect at its center. This solution can be derived from the stress field of a disclination, which obeys the equation  
\begin{align}\label{eq:app:fundumental_sol}
(\nabla^2)^2\chi = q \delta(\bs r)\,,
\end{align}
where the stress is related to the scalar field $\chi$ via a second derivative of the form 
\begin{align}\label{eq:app:stress}
\s^{ik} = \e^{ij}\e^{kl}\partial_{j}\partial_{l} \chi\,.
\end{align}
Then strain is related to stress in the standard manner 
\begin{align}\label{eq:app:strain}
\ve_{ij} = A_{ijkl} \s^{kl}. 
\end{align}
From equation Eq.~\eqref{eq:app:fundumental_sol} we obtain the fundamental solution of a dislocation by noting that a dislocation is a dipole of disclinations. Assuming the dislocation is a dipole along the $x$-axis we obtain such an expression using 
$$
\chi' = {b\over q}\partial_x \chi\,.
$$

The solution of Eq.~\eqref{eq:app:fundumental_sol} is given by 
\begin{align}
    \chi(r,\g) = c_1 \log r+c_2 r^2+c_3  +{qr^2 \over 8\pi} (\log r-1/2) 
\end{align}
Taking the derivative with respect to $x$ we then obtain 
\begin{align}\label{eq:app:stress_sol}
    \chi'(r,\g) = \left[{\tilde c_1\over r}+\tilde c_2 r  +{br \over 4\pi} \log r \right]\cos \g
\end{align}
Using Eq.~\eqref{eq:app:stress} we can then obtain the stress tensor. Finally, the values of the constants $\tilde c_1, \tilde c_2$ are dictated by the boundary conditions that require a vanishing stress normal to the boundary $\bs \s\cdot \bs n = 0$.

 \begin{figure}[]
	\centering
	\includegraphics[width=0.95\linewidth]{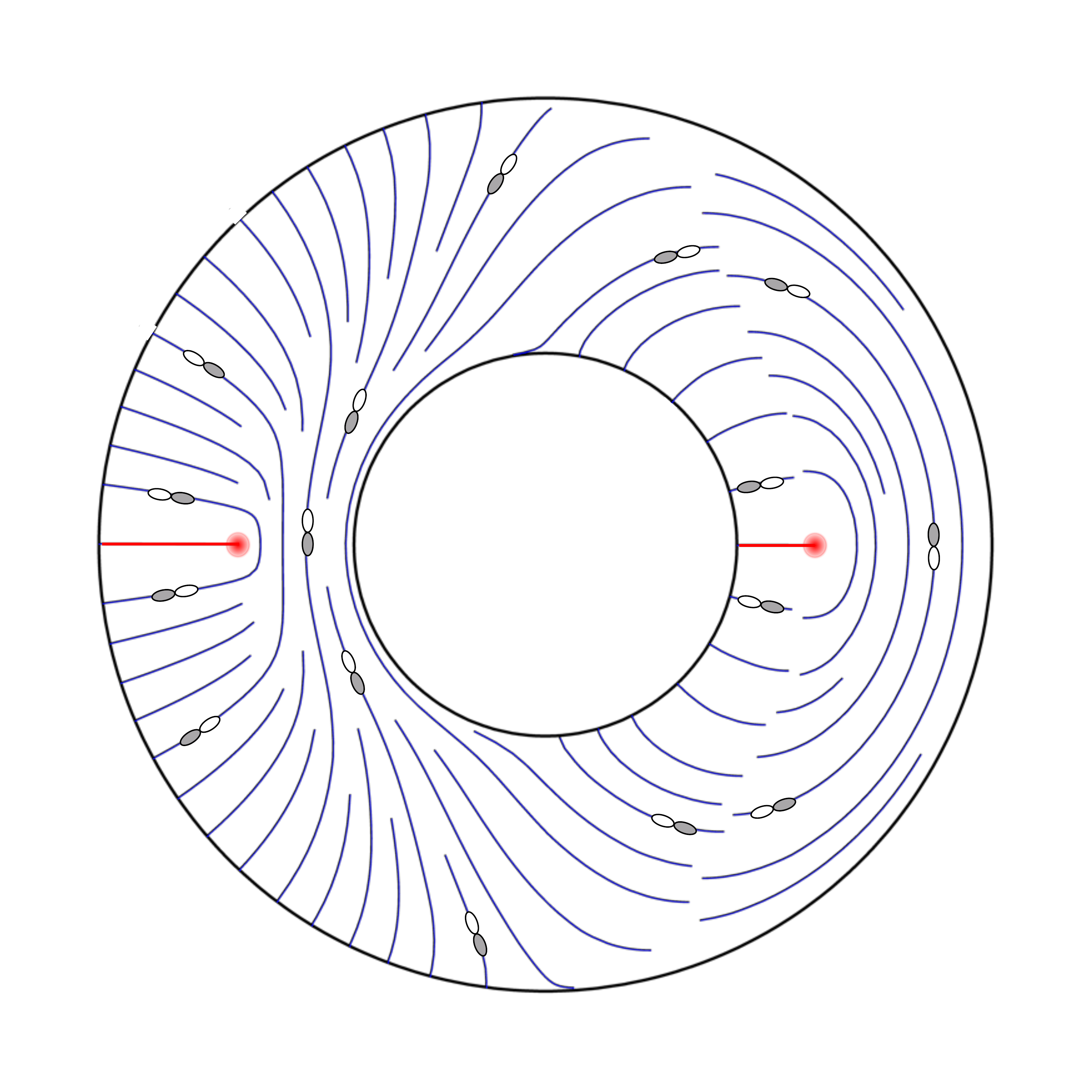}
	\caption{Strain field obtained from Eq.~\eqref{eq:app:stress_sol} of a two-dimensional ring with inner and outer radii ratio $R_1 / R_2 = 0.42$. Here a dislocation of arbitrary strength is assumed to be at the origin. The boundary conditions forces the strain field to develop two additional half-integer vortices located along the $y = 0$.  axes (red dots). Upon attempting to cover the strain field with a "nematic" p-wave superconducting order parameter one inevitably encounters branch cuts (marked in red), where the vector order parameter must vanish. Thus, such branch cuts will introduce $\pi$-junctions.    }
	\label{fig:supp:strain}
\end{figure}

The resulting strain field is plotted in Fig.~\ref{fig:supp:strain}. Here we used a "fat" ring  to make the strain texture visible. The topology in a thin ring remains the same.  As can be seen, the texture is more complex than the one depicted in Fig.~\ref{Fig3}. The boundary conditions induce two additional topological defects (red dots). 
It is evident that their topological charge is also halved. Indeed, any contour going around the hole and only one of the defects has an even number of branch cuts. 

A suggested covering of this strain field with a two-component (real)  order parameter is presented. The red lines mark branch cuts across which the order parameter field can not be glued and must change sign. As a consequence it will vanish near this region and develop a spontaneous $\pi$ junction.  When a $\pi$ flux is introduced to the center hole this frustration is removed, thus causing $T_c$ to increase. This will manifest itself as a $\pi$ shift in the Little-Parks oscillations.

% The structure is more complex than the one depicted in Fig.~\ref{Fig3}. However, the essential ingredient is the existence of an odd number of half-integer vortices in the strain field. This is a direct consequence of the assumption that a finite dislocation charge is trapped in the ring. Assuming the order parameter wants to minimize the number of such  $\pi$-junctions it will only have an odd number of them when an odd number of such defects is trapped in the ring. 

\subsection{Topological defects in strain and their relevance to the Little-Parks experiment}
The fact that $\pi$-junctions appear in certain samples regardless of temperature cycles above the   CDW transition is a key experimental observation. It implies the origin of $\pi$ junction is very likely a structural one. A second key observation is that the $\pi$ junctions are quite common (4 our of 9 samples exhibit this phenomena).

In the main text, we have argued that strain fields in the sample can align the two-component order parameter close to $T_{\rm c}$ via the $\kappa$-term in the GL free energy density Eq.~\eqref{eq:app GL theory}. In this way, the strain, which is embedded into the configuration of the order parameter, can frustrate the order parameter and force it to develop a $\pi$-junction~\cite{Geshkenbein1987vortices}, similar to the substrates in Ref.~\cite{Tsuei1994pairing}, which fixed the different crystallographic axis. Indeed, a two-component order parameter has been shown to be very sensitive to strain~\cite{Hicks}. Moreover, strain is expected to be a sample dependent feature explaining the stability of the $\pi$-flux in certain rings. 

\begin{figure}[]
	\centering
	\includegraphics[width=0.7\linewidth]{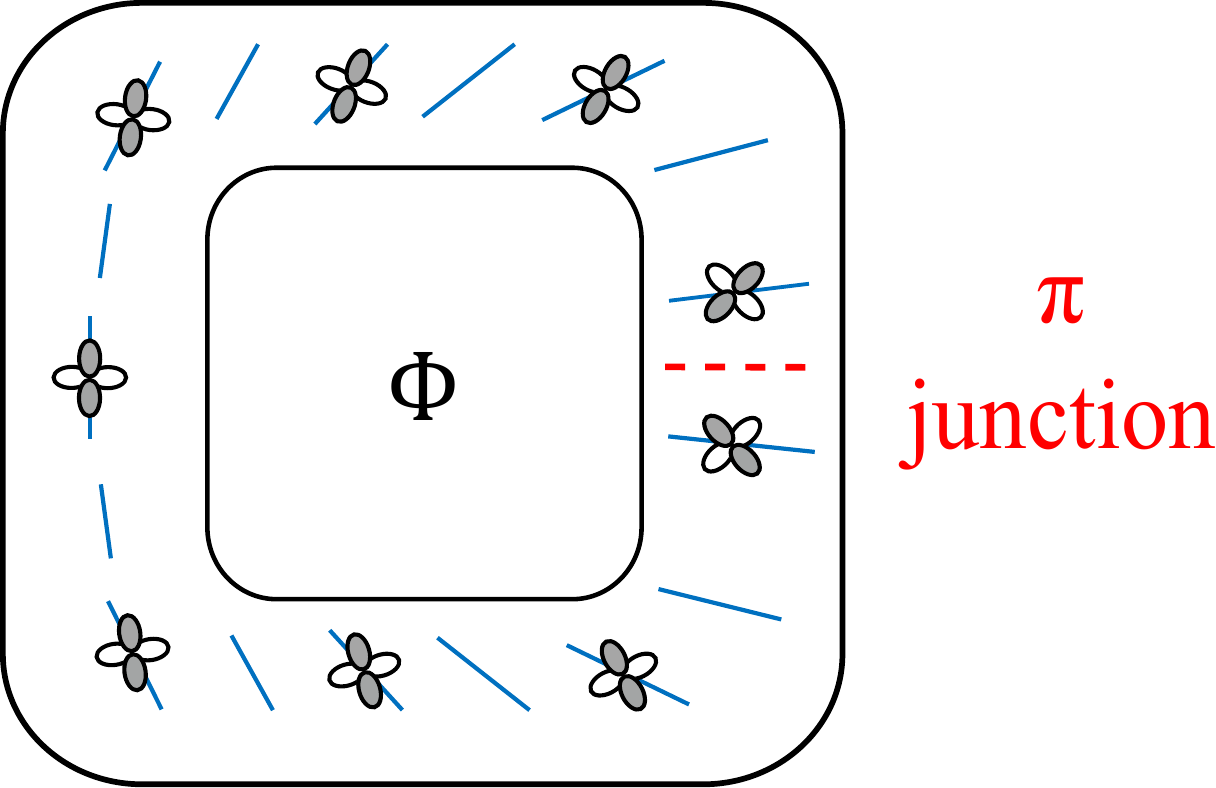}
	\caption{The formation of a $\pi$ junction for a ``d-wave'' ($E_{2g}$) order parameter. Note that the coupling to strain has the same form as for the $E_{1u}$ order parameter. In this case, the order parameter only ``rotates by $\pi/4$'' when the strain rotates by $\pi/2$.}
	\label{fig:d-wave}
\end{figure}

The question that remains is thus, what kind of strain field is required to force a $\pi$ junction and what can lead to such a field in half of our samples? The simplest scenario is depicted in Fig.~\ref{Fig3}. (For completeness we also schematically present the $\pi$-junction formation for a $E_{2g}$ order parameter in Fig.~\ref{fig:d-wave}.)  The strain, which has the structure of a headless vector can rotate by $\pi$ around the ring, without causing any inconsistency (like a disclination in a nematic medium). Unlike strain, the order parameter  is not compatible with such a rotation and is forced to go to zero somewhere in the ring, where a $\pi$ junction is naturally formed. 

A more realistic scenario is depicted in Fig.~\ref{fig:supp:strain}. The order parameter is forced to vanish and flip sign over a non-trivial contour in the ring. The exact form of the Little-Parks oscillation in the presence of such a strain field is beyond the scope of the current paper. However, the essential condition exists: The existence of $\pi$ junctions which can cause a $\pi$ shift in the Little-Parks oscillations.  

To show this we consider a semi-classical contour going around the inner part of the ring (ie for radii smaller than the two induced topological defects). In this case, the free energy is given by  
\begin{align}
    \delta f \sim -\kappa [ (|\eta_1|^2 - |\eta_2|^2)\cos \g  
    + (\eta_1^* \eta_2+ \eta_2^* \eta_1)\sin \g ]\,,
\end{align}
where $\g$ is the azimuth. 
Close to $T_{\rm c}$, this term will locally affect $T_{\rm c}$. The order-parameter configuration with highest transition temperature is then 
\begin{align}
    \bs \eta = \eta_0 (\cos{\g\over 2}\hat{\bs x} + \sin{\g\over 2}\hat{\bs y})\,.
\end{align}
However, such a configuration is obviously not compatible with a single-valued wave function. As a consequence the order parameter goes to zero across the branch cut, generating a $\pi$ Josephson junction. The free energy of the junction can be characterized by 
\begin{align}
    f_{\rm JJ} = E_{\rm J} \cos{2
    \pi \Phi \over  \phi_0} \,.
\end{align}


%\renewcommand*{\thefigure}{S\arabic{figure}}

%\end{figure}

\renewcommand*{\thefigure}{S\arabic{figure}}

\end{document}